\documentclass[11pt,a4paper]{article} 
\usepackage{amsfonts,amssymb,amsmath,graphicx} \usepackage{bm} \usepackage[spanish,english]{babel}
\usepackage[latin1]{inputenc}
\usepackage{hyperref}

\begin{document}

\title{``All that matter ... in one Big Bang ...,'' \\ \& other cosmological singularities\footnote{Parts of this paper have been presented by the author, under different titles, in the following 2017 Conferences, Seminars, and Workshops: (i) Universit\`{a} degli Studi di Trento, Trento, Italy; (ii) Kobayashi-Maskawa Institute for the Origin of Particles and the Universe, Nagoya University, Japan; (iii) Yukawa Institute for Theoretical Physics, Kyoto University, Japan; (iv) 5th CORE-U, Core of Research for the Energetic Universe, Hiroshima University, Japan; (v) 4th Korean-Japan Joint Workshop on Dark Energy at KMI, Nagoya, Japan; (vi) IV Cosmology and the Quantum Vacuum, Segovia, Spain; (vii) 3rd International Winter School-Seminar on Gravity, Astrophysics and Cosmology ``Petrov School'', Kazan Federal University, Kazan, Russian Federation.}
}

\author{{\large Emilio Elizalde} $^{1,2,3,\dagger}$ 
	\\ {\small $^1$  Spanish National Higher Council for Scientific Research, ICE/CSIC and IEEC} \\
{\small Campus UAB, C/ Can Magrans s/n, 08193 Bellaterra (Barcelona) Spain} \\
{\small $^2$  International Laboratory for Theoretical Cosmology, TUSUR University, 634050 Tomsk, Russia} \\
{\small $^3$  Kobayashi-Maskawa Institute, Nagoya University, Nagoya 464-8602, Japan }
  \\ {\small e-mail: elizalde@ieec.uab.es} \\
}

\maketitle

\begin{abstract}
The first part of this paper contains a brief description of the beginnings of modern cosmology, which, the author will argue, was most likely born in the Year 1912. Some of the pieces of evidence presented here have emerged from recent research in the history of science, and are not usually shared with the general audiences in popular science books. In special, the issue of the correct formulation of the original Big Bang concept, according to the precise words of Fred Hoyle, is discussed. Too often, this point is very deficiently explained (when not just misleadingly) in most of the available generalist literature.  Other frequent uses of the same words, Big Bang, as to name the initial singularity of the cosmos, and also whole cosmological models, are then addressed, as evolutions of its original meaning. Quantum and inflationary additions to the celebrated singularity theorems by Penrose, Geroch, Hawking and others led to subsequent results by Borde, Guth and Vilenkin. And corresponding corrections to the Einstein field equations have originated, in particular, $R^2$, $f(R)$, and scalar-tensor gravities, giving rise to a plethora of new singularities. For completeness, an updated table with a classification of the same is given.
\end{abstract}





\section{Introduction}

`Big Bang' is one among the very few scientific terms that have transcended its original domain of use and meaning, to become a common expression employed by literally everybody nowadays. It even gives name to a very popular TV series, and this is actually the first meaning that pops up now when you do a search on the internet. Many people associate Big Bang to an extraordinarily huge explosion at the beginning of everything, at the origin itself of our Universe, a bang which scattered all existing matter and energy that was concentrated in a nutshell just a fraction of a second before. Needless to say, this description, which at first sight might seem perfectly reasonable, makes very little sense, more than that, it is in fact utterly wrong, according to what cosmologists now know. But if we try to be more precise and resort to more specialized literature, there still is a lot of confusion going on, what is much more deceiving; the reason being that, commonly, the same two words are applied in several different contexts, with different meanings, never bothering to specify which is which. Only true specialists will not be lost, since they can easily distinguish the different contexts; but we cannot be satisfied with that, since the Big Bang issue has transcended the scientists' domains and the confusion will persists if we leave things as they are right now. To summarize, this concept  is in need of a lot of clarification, one should try to be more precise and clearly distinguish among these several versions and their exact meanings, in each situation, as we will now discuss.

There is, to start, the original meaning, the idea that Fred Hoyle intended to convey when he pronounced the two magic words in a BBC radio emission in 1949, giving them a meaning that was completely different from the one they had had before among cosmologists. This was
 in order to make an important point clear, concerning two competing models of the cosmos. Then, there is a second, more modern meaning, mathematically very precise (although physically much more blurred): the Big Bang singularity at the very origin of space and time. And there is also the concept of the Big Bang model, with two essential variants, the hot and the cold models, with hot or cold dark matter, variants that were under passionate discussion during several decades and until at least some twenty years ago, in the fine details.

To repeat, the author is quite convinced that these issues, although well-known to the true specialists, are in need of a serious discussion and clarification, what, unfortunately, can hardly be done in the reduced space of this article. But he will try, at least, to sketch and start walking along a possible way to be followed in further depth, in future analysis. To be remarked is that there are recent and very important discoveries by historian of science on different aspects of this subject, which need to be digested and put together into a global and detailed description, with the aim to draw a much more clear picture, replacing the wrong or misleading ones that commonly appear in popular references such as the Wikipedia, National Geographic, or even the British Encyclopedia. Maybe the main common problem has  always been, and is still now, the following: it turns out that the physics involved in a well-grounded explanation of those issues, namely General Relativity, is far from being trivial, more specifically, far beyond the proper comprehension of most of the authors of those chronicles and short descriptions. And, as we know well, a deep understanding is necessary before you are able to explain physics to your barber.

Going into the contents of this paper, after a brief approach, in Sect. 2, to the origins of modern cosmology, which the author will strongly argue it was born in the Year 1912, in Sect. 3 the first of the three different meanings of Big Bang, as discussed above, namely the original one, made explicit in the exact words of Fred Hoyle, will be addressed. Then, in Sect. 4, some of the main developments pertaining to the second meaning, the Big Bang singularity, will be summarized, starting with the classical singularity theorems. Concerning this same issue, quantum and inflationary additions to the celebrated singularity theorems by Penrose, Geroch, Hawking and others led later to two subsequent theorems by Borde, Guth and Vilenkin. And corresponding corrections to the Einstein field equations \cite{eins1} have originated, among others, $R^2$, $f(R)$ and scalar-tensor gravities, giving rise to a bunch of new (mainly future) singularities, which will be summarized there. Sect. 4 ends with a subsection that contains a short discussion of the hot and the cold Big Bang models (the third distinct meaning of these words), supplemented with a selected bibliography of the same. For completeness, in Sect. 5 a table with the classification of the new cosmological singularities, which show up in theories of modified gravity and are now under hot discussion, in the context of the accelerating universe, is given. Finally, Sect. 6 is devoted to conclusions.

\section{The very origins: Leavitt, Slipher, Hubble, Lema\^{\i}tre, Einstein, ...}

This section is devoted to a short remebrance of the origins of modern cosmology. Common to the study of any dynamical system, the two main issues in the discussion of the dynamics of the cosmos are the determination both of the distances to celestial bodies, and of their velocities. But it turns out that calculating distances to far celestial objects has always been, and is still, one of the most difficult tasks in astronomy. The reader will surely remember the enduring discussions, which lasted for several years, concerning the supernovae SNIa as standard candles (the possible role of dust, of large-scale matter inhomogeneities, etc.), in the precise definition and evolution of the distances with time, which eventually led to the highly surprising and remarkable result of the acceleration of the cosmic expansion.

\subsection{Old models of the Universe}

But here we are looking back into the early history of modern cosmology, which started, in the authors' opinion, in the first half of the second decade of last Century. Before that, however, we should throw a quick look into a much more remote past, with the main purpose of realizing how difficult it is to calculate astronomical distances. Recall that in the first models of our Universe, as the celebrated one by Anaximander (ca. 610 - 546 BC) \cite{anax1} an early pre-Socratic philosopher from the Greek city of Miletus,  the Sun was considered to be the most distant object from the Earth; and this in spite that, in this time, Greeks could see in the night sky many more stars and galaxies with the naked eye that we are able to observe from our polluted cities nowadays. Of course there was a scientific reason for the Sun being that distant, since (recall the four-element theory) the Sun is fire and fire always goes up, therefore... In Anaximander's universe the Moon was the second most distant object (another fire, but smaller than the Sun), while stars were constrained to a cylinder much more close to us. Anaximander's universe was very advanced to his time; apparently he was the first cosmologist to get rid of Atlas, the hero who prevented the Earth from falling down and which had been omnipresent in  previous Greek models of the cosmos. For Anaximander everything, including the Earth, was floating in space (the ether, if you want) in perfect equilibrium, with the measures of the Earth, the radius of the Moon and Suns orbits, etc., keeping accurate harmonic proportions.

In the Ptolomean universe of the second Century AD there are several important changes, which for the sake of conciseness will not be discussed in any detail here \cite{aheld1}. Although for Ptolemy the stars were already further from the Sun (at distances from the Earth of 20,000 vs 1,200 Earth radii, respectively \cite{akw1}), it is not until the Copernican revolution that we can find an `official' map of the cosmos where stars are depicted outside and far away filling the cosmos, the Sun lying now at the center of the universe, as can be seen in the very famous Thomas Digges' representation, of 1576, of the Copernican universe \cite{harris1}. This was a true revolution that went on during the centuries to follow, to eventually `enlarge' the dimensions of the observable Universe to those of our Milky Way. And here ends our short travel to the remote past.

Such was actually the situation at the beginning of the XXth Century. Indeed, the Milky Way was the entire observed Universe: all stars, spiral nebulae, and other celestial objects were considered to be inside of our own galaxy, none of them was suspected to lie far beyond its boundaries. Moreover, all those objects were moving in perfect equilibrium, that is, the Universe was static. Again, there was a strong scientific reason for that, for such is the final result of the evolution of any physical system under quite general conditions and the Universe, being eternal, had had enough time to evolve and reach such regime. It was precisely this beautiful model of the Universe: eternal, static, and small as the Milky Way, which led Albert Einstein to the most horrible blunder of his whole life: {\em ``... die gr\"{o}{\ss}te Eselei meines Lebens!''} where, according to George Gamow, his exact words \cite{gamow1}.

\subsection{Henrietta Leavitt}

By now we should have grasped already how extremely difficult it is to calculate distances in astronomy. The order of magnitude of the mistakes committed in this respect during past epochs are just colossal. And this is precisely why the first hero in our story, in fact a she-hero, a heroine, must undoubtedly be Henrietta Swan Leavitt. Such was the importance of the extraordinary discovery she did in 1912 --after several years of collecting thousands of data, in particular from the Magellanic clouds-- namely the period-luminosity relationship of Cepheid variable stars: a linear dependence of the luminosity vs the logarithm of the period of variability of the star's luminosity \cite{leavi1}. It would be interesting to describe the favorite physical mechanisms available to explain such relationship (as the Eddington valve \cite{edval1}, for a very beautiful one), but regretfully there is no place here for that. Henrietta was a distinguished member of the so-called Edward Pickering's Harvard harem, or better known as Harvard computers, a group of young ladies that did a tremendous job in astronomy during that time (the interested reader can find in, e.g. \cite{leavi2}, more details). Leavitt's result was an extremely powerful weapon to calculate distances, in fact the main tool used by Hubble in the  years to follow, and subsequently by several generations of astronomers with remarkable success, until the advent of other improved techniques \cite{rr1} that have culminated in the SNIa standardizable candles mentioned above \cite{stcan1}.

\subsection{Vesto Slipher}

Our second hero in this story is, again by his extraordinary merit, the well-known astronomer Vesto Slipher who, starting on the very same year 1912 obtained the first radial velocities of spiral nebulae from their spectral blue- or red-shifts, by using the 24-inch telescope of the Lowell Observatory, in Arizona. Actually his  first calculation, which he produced on Sep. 17, 1912, was for the Andromeda nebula, a blueshift  \cite{sliph1}. In 1914, in a meeting of the American Astronomical Society, he presented results for 15 nebulae,  results that were received by the audience (chronicles say) with a very long, standing ovation  \cite{sliph2}. This is most unusual, in a scientific conference, then and now. He was, without any doubt (as Hubble himself later recognized), the first astronomer to see that something highly remarkable and very strange was going on in the static model of the cosmos: those distant nebulae escaping from us at such enormous speeds, and it is very clear that the importance of his discovery was immediately appreciated by the attendees, according to the chronicles\cite{sliph2a}.

Now, with {\it distances} and {\it speeds} the two necessary tools were ready for the greatest revolution in the study of the cosmos to occur: a radical conceptual change that marked the beginning of modern cosmology.

\subsection{A Great Debate}

Before continuing along this line, a very short mention of the so-called Great Debate \cite{gdeb1}, Harlow Shapley vs Heber Curtis, which took place at the Smithsonian Institution in Washington on Apr. 26, 1920. Shapley defended the orthodox view that the Milky Way was the entire Universe, while Curtis, his opponent, raised serious doubts, one of his arguments being that an unusual concentration of novae stars had been reported in Andromeda, this pointing to the possibility that such nebulae was in itself another world, another universe disconnected from us and similar  to the Milky Way.

In any case, it seems clear from the chronicles  that, in this case, nobody won that debate, none of the two astronomers could convince the opponent with his arguments and, at the end of the day, the result of the scientific contest was a draw. There are several very nice accounts of this famous debate and of this epoch of astronomy (see, e.g., \cite{mbart1}, and for additional material on subsequent developments \cite{dover1}).

\subsection{An Island Universe}

Nov. 23, 1924, is the following important date in this story. That day, on the 6th page of the New York Times the following  news appeared:  \begin{quotation}  ``FINDS SPIRAL NEBULAE ARE STELLAR SYSTEMS; Dr. Hubbell Confirms View That They are `Island Universes' Similar To Our Own.

WASHINGTON, Nov. 22. Confirmation of the view that the spiral nebulae, which appear in the heavens as whirling clouds, are in reality distant stellar systems, or ``island universes,'' has been obtained by Dr. Edwin Hubbell of the Carnegie
Institution's Mount Wilson observatory, through investigations carried out with the observatory's powerful telescopes.''  \end{quotation}
Some months before that day, in 1923, Edwin Hubble had spotted a Cepheid variable star  in a (now very famous) photographic plate of the Andromeda nebula, namely the variable Cepheid star V1  \cite{cefpl1}  and, using Leawitt's law, he was able to calculate a distance of 285 kpc to the Cepheid. To his surprise, this was about ten times larger than any other distance calculated before for celestial objects in the Universe, what very clearly supported the view that Andromeda was definitely outside of the Milky Way. Hubble presented his results to the Meeting of the American Astronomical Society that took place in Washington starting Dec. 30, 1924. On Jan. 1, 1925, he submitted his contribution, which was read in fact by Henry Russell, the director of the Princeton University Observatory, who had forced Hubble to present his still unfinished paper. Its title was {\em ``Cepheid variables in spiral nebulae''}, and was read in the joint morning session of astronomers, physicists and mathematicians. It won the year's 1,000 dollar prize of the American Association for the Advancement of Science (to concur to the prize was the real reason for such urgency).

This was indeed an astonishing discovery, one that finally decided the result of the Great Debate, substantiated the ideas of philosophers, as Immanuel Kant, and completely changed our vision of the cosmos. It has been righteously highlighted in innumerable occasions.

However, what Hubble seems not to have been aware of at that point, and many scientists (and, incredibly, most of the references to the subject) still ignore {\it even today!}\footnote{This being one of the reason why it is my inescapable duty to explain all that here with such emphasis.} is that, two years before Hubble, a well-known Estonian astronomer, Ernst \"{O}pik, had published a paper in the prestigious Astrophysical Journal \cite{opik1}, in which he had obtained a distance to Andromeda of 450 kpc, what is much closer to the actual value of 775 kpc than the value obtained by Hubble! \"{O}pik used a very different method to calculate the distance \cite{opik1}, which has nothing to do with the presence of a Cepheid star. His method was based on the observed rotational velocities of the galaxy, and on the assumption that the luminosity per unit mass was the same as that of the Milky Way. Although \"{O}pik was a reputed astronomer at the time and, on top of this discovery, he was the first to calculate the density of a white dwarf, very few, outside of the community of astronomers, remember his name now.

\subsection{Hubble's law}

In the late 1920s, putting together the table of {\em Radial velocities in km/s  of 25 spiral nebulae} published by V.M. Slipher in 1917 \cite{sliph3} (which, by the time, had even appeared in Eddington's book \cite{eddi1}) and his own table of {\em Distances in Mpc of spiral nebulae}  \cite{hubb1}, Edwin Hubble obtained the very famous law that bears his name, and published it in 1929 \cite{hubb2}. When one puts side by side the two tables one immediately realizes how easy is to fit the values to a straight line, $V = H_0 D$, with $H_0$ a constant, named now after Hubble. He did not mention in his paper that Slipher was the author of the redshift table and, even today,  in many references to his work he is wrongly considered to have produced both tables: the one of the distances and the other of the redshifts.

It is fair to say that Hubble did later recognize Slipher's remarkable contribution, in all of its importance. Indeed, in a  Letter  to V.M. Slipher, of March 6, 1953 \cite{husli1},  he wrote: \begin{quotation}  {\em ``... your velocities and my distances...''} \end{quotation}

More than this, Hubble acknowledged the great influence of Slipher's seminal and important contribution to his own subsequent work by declaring that \cite{husli2}:  \begin{quotation}
{\em ``... the first steps in a new field are the most difficult and the most significant. Once the barrier is forced further development is relatively simple.''} \end{quotation}
 In fact, it was Slipher the first to realize, as we have explained above, that something very important and remarkably strange was going on in the static Universe model: how on Earth could it be static with those distant nebulae getting away from us at such enormous speeds?

\subsection{The interpretation of Hubble's law}

As advanced, Hubble's law is quite easy to obtain from the two aforementioned tables, but this is by no means the whole point, not even the {\em main} one. The key issue is the {\em interpretation} of Hubble's law. Namely, (i) do the high escape velocities of the spiral nebulae correspond to {\em real} displacements of the celestial objects (as was believed by {\it everybody} at the beginning) or, (ii) on the contrary, they correspond to the movement of the {\em reference system}, to an expanding space?  Of course, the obvious answer is that, in fact, both contribute to the redshift; even today this is a most difficult problem in astronomy: to disentangle these two components in the observational redshift maps of astronomical objects. But we are here talking of the movement of very distant objects where we know now that the second interpretation (and the corresponding contribution) prevails, without any doubt. This interpretation was extremely difficult to understand at the beginning and the only accepted explanation was the first (what no popular-science writer seems to be able to grasp now, in retrospect!).

Further to this point. From recent studies of historians of cosmology it seems now clear that Hubble never believed that the universe was expanding. What is without question is that he never wrote this statement in any of his works. In a letter of Hubble to Willem De Sitter, in 1931, he stated his thoughts about the velocities by saying: \cite{hubb3} \begin{quotation} {\it
``... we use the term `apparent velocities' in order to emphasize the empirical feature of the correlation. The interpretation, we feel, should be left to you and the very few others who are competent to discuss the matter with authority.''} \end{quotation}

A second important remark is also in order. It is written in books and in many other places in the literature that it was Hubble the one who convinced Einstein that the Universe was expanding, when the later visited Hubble at Mount Wilson in 1931, during his famous tour in the USA that year. But, examining in detail Einstein's notebooks and other writings, it has been unveiled now that he was actually convinced --in 1931 in fact-- by Eddington, Tolman, and de Sitter {\it (not at all by Hubble)} of both the fact that his static model was unstable and that the Universe was actually expanding  \cite{nuss1}.

As is well known, Einstein had introduced his famous cosmological constant  in 1917 (exactly 100 Years ago now), as an additional term to his field equations of General Relativity --which he had previously derived, in final form, in 1915-- in his attempt to describe with them the whole Universe.\footnote{Theoretical cosmologists may, alternatively, consider the Year 1915, when Einstein completed his formulation of the gravity field equations, incorporating his revolutionary principle of equivalence, as the one actually marking the beginning of modern cosmology. This is, in particular, the opinion of one of the referees of the present paper. Indeed, the moment Einstein had his ``most happy thought'' unveiling the relativistic role of the equivalence principle is another seminal event from which the whole theoretical framework stems. Also, the expansion of the universe was first apparent in the work of Friedmann, and partly on de Sitter's 1917 contributions. I am myself a theoretician and, although I recognize that GR is probably the most beautiful and transcendental theory of Physics ever conceived and constructed, I still stand by my opinion, as expressed above. Which is also supported by the fact that the very important developments of 1912 preceded those of 1915 and 1917.} As we have pointed out, the Universe, at that time, was considered to be static and everybody believed that it was reduced to the Milky Way. But a static Universe is not a solution of the original Einstein's equations (nor of Newtonian physics either), since it will definitely collapse under the influence of the gravitational force. With an appropriate sign, the cosmological constant would provide a repulsive force preventing this collapse. When Einstein finally got convinced of the Universe expansion, he pronounced his famous sentence recognizing his horrendous mistake, as reported above.

\subsection{George Lema\^{\i}tre and the expanding Universe: a perfect example of Stigler's law of eponymy}

But, who was the first person on Earth to understand that space, the fabric of the Universe was expanding? We come here to our next huge surprise: it turns out that Hubble was not the first to derive Hubble's law. There is a famous principle, widely known now under the name of Stigler's law of eponymy \cite{sle1}, which states that:  \begin{quotation} {\em ``No scientific discovery is named after its original discoverer''}. \end{quotation} The reader may have realized that we are encountering several clear examples of this principle here.

While working for his PhD Thesis at MIT (Cambridge, MA, USA), George Lema\^{\i}tre rediscovered Friedmann's mathematical solution to Einstein's equations \cite{fried1}. But he was in no way a mere mathematician, but a true physicist pursuing to build a cosmological model in accordance with the astronomical observations. No wonder then that he went to visit the most prominent astronomers of the time, in particular Vesto Slipher, at Lowell Observatory in Arizona, and Edwin Hubble at Mount Wilson, and got from them the two aforementioned tables, namely the one of redshifts, and that of distances. It was for him a child's play to discover the correlation and immediately obtain Hubble's law, two years before Hubble.\footnote{Andrzej Wr\'{o}blewski brings to my attention the two papers by Carl Wirtz  \cite{cwirt1}, of 1922-24, where he derived a linear relation of recession of spiral nebulae with their distance. Wirtz used the assumption that all galaxies have approximately the same size and estimated their relative distance from their apparent diameters. For unknown reasons, Wirtz's papers were not noticed or soon forgotten (although they are certainly mentioned by historians of astronomy).}

Actually, he did this only after having submitted his PhD Thesis at MIT in 1925, since he had had no time in Boston to complete the whole study leading to his cosmological model. What he finally finished on his return to Belgium, to teach at the Catholic University of Leuven. In 1927 he published his complete results, in a  Belgian journal of very low impact (Annals of the Scientific Society of Brussels) \cite{lemai1}. Those who know French may check that, in fact, Hubble's law is already there, with a value of the Hubble constant very close to the one obtained by Hubble in 1929, since the late just added a couple of extra nebulae to the tables (with the new redshifts having been obtained by his coworker Milton Humason) \cite{lemai2}.

During the Fifth Solvay International Conference on Electrons and Photons, which took place in Brussels in Oct. 1927 (a most famous meeting, because 17 of the 29 participants appearing in the celebrated picture of the meeting were already, or later became, Nobel Prize winners), Lema\^{\i}tre approached Einstein, handing him his recently published paper. He told Einstein, in short, that he had discovered a solution to his original field equations, which would correspond to an expanding universe, perfectly matching the latest astronomical observations. His cosmological constant was not necessary at all, he told Einstein, quite on the contrary, the static universe solution that Einstein obtained by adding it was, in fact, unstable! Some days later, after having examined Lema\^{\i}tre's calculations with some care, Einstein's answer to him was that he could not find any mistake in the mathematical formulas but that his physical insight, the fact that the universe was expanding, was nonsensical ({\em abominable}, in French, as Lema\^{\i}tre reported later).

It turned out that Einstein had already rejected this same idea in an answer to a letter from Friedmann on the same issue, some years before. Even worse, his first comment to Friedmann's work had been that his equations were in error! an objection that he had to retract few weeks later, after a complaint by Friedmann that this was not at all the case \cite{einsfr1}.  But even this had not convinced Einstein of the feasibility of an expanding universe. Nor was he convinced by Lema\^{\i}tre and by the clear astronomical evidence that he showed to him, this last for the first time.

The (rather astonishing) fact that it took Einstein --the creator of General Relativity, the master of space and time, the discoverer of gravitational lenses and gravitational waves-- still four full years, until 1931, to become convinced of the fact that the Universe was actually expanding may seem now very strange, even unbelievable. On the contrary, this just supports the author's conviction that such was indeed, at the time, an extremely revolutionary idea, which many (starting with Hubble) could not comprehend in all its significance.

Lema\^{\i}tre was, for a while, the only guy on Earth who was comfortable with the idea that space was in fact expanding, and he had a very hard time to convince other colleagues of this discovery. The first one to understand his model was Arthur Eddington, who knew Lema\^{\i}tre well from a visit of the later to Cambridge (England), previous to his stay at MIT.  Eddington, for one,  had also proven by himself that Einstein's static solution (of the equations with the cosmological constant) was unstable, and he was already working along the idea of a possibly expanding universe, so that when he saw Lema\^{\i}tre's paper he realized at once that this was the solution he had been looking for. He was positively surprised to see all the work already done, and even more, that the result was in full accordance with the astronomical data.

In 1930, Eddington published, in the Monthly Notices of the Royal Astronomical Society, a long commentary on Lema\^{\i}tre's 1927 article, in which he described it as a ``brilliant solution'' to the outstanding problems of cosmology \cite{edd1}. Moreover,
Eddington helped Lema\^{\i}tre to translate his paper to English, and it was published in March 1931 in the prestigious Monthly Notices of the Royal Astronomical Society, but only the first part ({\em ``premi\`{e}re partie''}) of it \cite{lemai3}, which does not contain Hubble's law. For some decades it remained a mystery why the second part of the paper was not translated. It now seems that this was not prevented intentionally by an anonymous referee, but that it actually was a personal decision of Lema\^{\i}tre's himself, who did not consider these results to be so important any more, after the appearance of Hubble's paper that improved them somehow \cite{lemai2}.  I would add to this the comments to be found in Ref. \cite{gmm1}, and my own personal considerations, which I think are quite reasonable: Lema\^{\i}tre must have been perfectly aware of the fact that the tables of data from which he had obtained Hubble's law before Hubble himself were handed to him, graciously, by their authors, namely Slipher and Hubble. To derive from them the correlation had been a simple exercise.

However, it still remains, as stressed above, the most crucial issue of the interpretation of the law as an expansion of the Universe, and in this aspect Lema\^{\i}tre has no rival; quite on the contrary, he was so much ahead of everybody else that he had a very hard time trying to convince the rest of cosmologists that this was the true reality, indeed. Anyhow, that I know, Lema\^{\i}tre never complained, in his whole life, for not having been credited with the discovery of the Universe expansion.

\section{``All matter ... in one Big Bang ...''}

In this section we will explain the original meaning of `Big Bang', as crystal clearly expressed in the sentence of Fred Hoyle, the man who  pronounced these two words together, with a brand new signification, in 1949.\footnote{After completion of version 1 of the present paper, I got through a very interesting preprint \cite{kragh1}, which presents a nice account of names and concepts associated with finite-age cosmological models from the 1920s to the 1970s, including the many meanings of the Big Bang name. This reference is quite complementary to what I am describing here; it does not go however, in any detail, into the most crucial point of the  actual meaning of a Big Bang in the Theory of General Relativity, which is the essential issue in this section.}

Looking backwards in time, into the past evolution of the cosmos, Lema\^{\i}tre judiciously argued that if the Universe was expanding it ought to have had an origin.\footnote{To be mentioned is that there were at the time other competing models of the cosmos, as e.g. the interesting contribution of H.P. Robertson \cite{roberts1}.}  That is, at the beginning, all matter and energy, as space itself, should have been constrained to a small region, a nutshell. In a meeting of the British Association on the relation between the physical universe and spirituality, he proposed in fact that the Universe expanded from an initial point-like structure, namely a {\em ``Primeval Atom''}, or a {\em ``Cosmic Egg exploding at the moment of the creation``}, as his precise words were. In 1931, he published this theory in Nature \cite{lemai4}. It is really shocking that this erroneous, absolutely misleading (in our present understanding) theory of the origin of the Universe is now much better known to the public than Lema\^{\i}tre's extraordinary  contributions and insight concerning the expansion of the cosmos, as described in detail in the previous section, and which are largely ignored, all tribute going to Hubble and Friedmann. Maybe the reason is that everybody understands what an explosion is, sending debris everywhere, even if it is that of an enormously huge bomb. But who is really able to grasp the sense of the fabric of space expanding extremely fast?  and thereby allowing for the possibility (according to GR) of the creation of big amounts of matter and energy (e.g., eventually, the quark-gluon plasma) {\em out of nothing!}. Moreover, that the total content of matter and energy of the Universe is zero (or almost zero, for all we now know). Nobody, not even a vast  majority of trained physicists, much less popular science writers (as I will certify below), but only true specialists on the subject of GR can actually deal with these concepts.

To wit, I will here reproduce some of the crazy definitions of ``Big Bang'' recently encountered in the internet, in different places and languages.
\medskip

\subsection{Some common popular sources on the Big Bang} \medskip

\noindent {\bf  Wikipedia webpage:} ``If the known laws of physics are extrapolated to the highest density regime, the result is a singularity ...'' ``Since Georges Lema\^{\i}tre first noted in 1927 that an expanding universe could be traced back in time to an originating single point, scientists have built on his idea ...'' ``Extrapolation of the expansion of the universe backwards in time using general relativity yields an infinite density and temperature at a finite time in the past ...''
\medskip

\noindent {\bf French Wikipedia:} ``De fa\c{c}on g\'{e}n\'{e}rale, le terme ``Big Bang" est associ\'{e} \`{a} toutes les th\'{e}ories qui d\'{e}crivent notre Univers comme issu d'une dilatation rapide qui fait penser \`{a} une explosion ...''
\medskip

\noindent {\bf Italian version:} ``La fase iniziale calda e densa \`{e} denominata ``Big Bang'' ...''
\medskip

\noindent {\bf The National Geographic:} ``Before the big bang, scientists believe, the entire vastness of the observable universe, including all of its matter and radiation, was compressed into a hot, dense mass just a few millimeters across.''
\medskip

\noindent {\bf Global Britannica:} ``Its essential feature is the emergence of the universe from a state of extremely high temperature and density: the so-called big bang ...''
\medskip

\noindent {\bf NASA webpage:} ``Was the Big Bang an explosion?  No, the Big Bang was not an explosion. We don't know what, exactly, happened in the earliest times, but it was not an explosion in the usual way that people picture explosions. There was not a bunch of debris that sprang out, whizzing out into the surrounding space. In fact, there was no surrounding space.  There was no debris strewn outwards. Space itself has been stretching and carrying material with it.''
\medskip

Of all those, only the very last definition, and in part the last but one, in its indefiniteness, can be saved. All the rest reduce to re-formulations of Lema\^{\i}tre's concept of the (in his own words) {\em ``Cosmic Egg exploding at the moment of the creation``} that was proven to be absolutely erroneous and fully misleading by nuclear physicists already before the end of the thirties of last Century \cite{addu1}, and onwards to the 1940s \cite{nphy1}, where the activity in this direction was extraordinary. For many reasons, it was by then already appreciated that it was absolutely impossible that the whole matter and energy of the present universe could have been initially present, already, and confined to a nutshell. What is most incredible is that, if this impossibility was so crystal clear almost 80 years ago, this utterly wrong definition continues to be present in almost all books, encyclopedias and general references today.

\subsection{Fred Hoyle}

Further to the point, among these nuclear physicists and astrophysicists there was one named Fred Hoyle (1915-2001), an English astronomer noted primarily for the theory of stellar nucleosynthesis, on what he wrote groundbreaking papers \cite{fhoy1}. During the II World War, he worked on Britain's radar project with Hermann Bondi and Thomas Gold. Although Hoyle never got the Nobel Prize, his colleague William Fowler, who did get it in 1983 for work on stellar nucleosynthesis, recognized that: \begin{quotation} {\em ``The concept of nucleosynthesis in stars was first established by Hoyle in 1946''}. \end{quotation}  Hoyle had always found the idea that the universe could have a beginning to be pseudoscience, mere arguments for a creator, \begin{quotation} {\em ``... for it's an irrational process, and can't be described in scientific terms''; ``... belief in the first page of Genesis''}. \end{quotation}  Hoyle, Gold and Bondi published in 1948 their (later quite famous) {\em steady state theory}, involving a {\em ``creation or C-field''}. The reasoning was the following. Like many other physicists in the 1940's, they continued to believe that the static model of the Universe was the right one. There had never been any doubt that Hubble's law was correct; therefore, in order to compensate for the matter density loss due to the distant galaxies going away from us, they had to introduce a term in their equations, which created matter in far regions of the cosmos, at a rather smooth rate. This was the creation or C-field. And how did they manage to generate matter out of `nothing'? Very simple, they just used General Relativity to do this job. They involved in their theory exactly the same physical principle that allows for the creation of the quark-gluon plasma in most of present day's inflationary theories.

\subsection{The `free lunch' concept}

Regretfully, nowadays it is not so widely known that the concepts of a universe of {\em ``zero total energy"}, or the {\em ``free lunch"}, namely keeping this zero-energy balance along the formation and subsequent evolution of the universe, are not concepts invented by A. Guth, A. Linde or A. Vilenkin, nor by any of the inflation physicists. These concepts are very clearly explained already, e.g., in the famous book by Richard C. Tolman of 1934, {\em ``Relativity, Thermodynamics, and Cosmology"} \cite{tolm1}. There, one finds how \begin{quotation} {\em ``... a closed universe can equal zero energy. All mass/energy   is positive and all gravitational energy is negative, and they may cancel each other out, leading to a universe of zero energy''}. \end{quotation}  This is now called, in Allan Guth's brilliant lectures at MIT the {\em ``Miracle of Physics No. 2"} \cite{ee1}. It is preceded by the {\em ``Miracle of Physics No. 1"}  \cite{ee1}, as explained in the same lectures, which is actually the one that interests us now, to begin with. Keeping the energy balance (or principle of energy conservation) all the time, in accordance with General Relativity, and in particular with Friedmann's equations, where we see that matter/energy density goes together with pressure of the reference system, it turns out that a positive amount of matter/energy can be generated provided an equivalent amount of negative pressure, e.g. an expansion of the reference system (in mathematical terms), or of the fabric of space (in physical ones), is available (aka inflation).

But we have got too far along this way and we need now get back to the point. Hoyle perfectly new, in the late 1940s, that it was absolutely impossible for the whole matter and energy of the  universe to be initially present, already at the very beginning, and confined to a nutshell. To start with, all but the three or four lightest elements had to be generated under much more energetic conditions, e.g. in star explosions (stellar nucleosynthesis), a physics that he pioneered. And he also realized that, in Lema\^{\i}tre's model, these lightest elements had to be generated {\em all at once} at the beginning of the universe: all that huge amount of matter, and instantly! --unlike in his coauthored {\em steady state theory}, where this took place quite smoothly, in far apart regions of our universe, and in small proportions. In Lema\^{\i}tre's model one would need such an enormous blow of space, an incredibly large negative pressure of the reference system, in order to be able to create, instantly, such huge amount of positive matter/energy.

\subsection{A Big Bang!}

This is what Hoyle had in mind, and it is exactly, word by word, what he said in the now very famous BBC radio's Third Program broadcast of March 28th, 1949:
\begin{quotation}
 {\em ``[Lema\^{\i}tre's model implies that] ... all matter in the universe was created in one {\bf Big Bang} at a particular time ...''}.
\end{quotation}

This one (and no other!) is the precise meaning of `Big Bang'\footnote{John Barrow has told me \cite{jbarrpc1} he had always suspected that the term `Big Bang' had been remembered in Cambridge, before that episode, from its use by Eddington in his book {\em ``The Nature of the Physical World''} \cite{Edding2014a}, based on his famous Gifford Lectures of 1927 at the University of Edinburgh \cite{giff27a} (probably the most famous public lecture series in the world then). Eddington writes,  {\em ``As a scientist, I simply do not believe that the Universe began with a bang''} --see also \cite{jbarrb1},  pgs. 123 and 318. Also according to Barrow, Eddington had a more sophisticated view of the origin of the universe
than others, and he regarded the Eddington-Lema\^{\i}tre models, which are geometrically past eternal, as having finite thermodynamic age because only a finite number of non-equilibrium events would have happened to the past. } according to the person who came up with these two words in order to express the idea of an {\em impossible blow of space} being needed to create all that matter in the universe instantly, in accordance with the fundamental principles of General Relativity.

Now, again, how can one explain that such well formulated concept, rigorously expressed, with so precise words from the very beginning, is today explained in popular references in such absurd ways? In my view, I repeat, this is because while everybody understands the meaning of a Bang as an ordinary explosion scattering matter in all directions, only very, very few will understand the concept of a Bang of the fabric of space, an enormous blow that allows for the creation of large amounts of matter out of nothing, without violating the energy conservation principle.\footnote{In particular, very few realize the radical difference between a `Bang' and the `Big Bang' concept, in spite of the fact that (or maybe precisely because) everybody knows what `Big' means. That a Bang was necessary at the origin of the world was naturally understood by everybody accepting Hubble's law, and this included many cosmologists of the time (and a lot of educated people nowadays). In this precise way was it used in Cambridge (see the previous footnote, reflecting Barrow's interesting observation). But it turns out that `the Big Bang' in the sense of Hoyle is {\it not} just a Big Bang in the common sense, namely of a very huge Bang of the ordinary sort, giving the objects in the Universe a very large recession speed. The crucial point is that of matter creation (since the objects were not there! to start with), almost instantly, in a very brief inflationary process (as is called now) at the very origin of the cosmos. I hope this is clear already.}

I must stress again the enormous contradiction pervading all these popular references above. Lema\^{\i}tre is now remembered for the {\em wrong} reason, namely for his primordial atom model --that was so far from reality-- and never for his fabulous insight and wonderful cosmological model of 1927, so much advanced to his time that not even the greatest physicists of the epoch, Einstein included, could understand. Exactly the same misunderstanding happens with Hoyle, who is now only remembered as the proposer of the discredited Steady State theory of the universe, which proved to be a wrong model in the end --and for having prevented the teaching of the universe expansion in Cambridge during decades, even much after the Cosmic Microwave Background (CMB) radiation was detected \cite{cmb1} (also his panspermia ideas did not help at all). But this attitude was just because he considered that such an incredible blow of space, which he called Big Bang and it is now called inflation, and has become the standard theory of the universe origin, was absolutely impossible to happen, it could not be! in his understanding.

John Gribbin, in Hoyle's obituary, beautifully called {\em ``Stardust memories"} \cite{jgri1},  writes:
\begin{quotation}
{\em ``Everybody knows that the rival Big Bang theory won the battle of the cosmologies, but few (not even astronomers) appreciate that the mathematical formalism of the now-favored version of Big Bang, called inflation, is identical to Hoyle's version of the Steady State model''}.
\end{quotation}
Truly, Hoyle was the stardust guy, the man who proved to us that we are all stardust, our bodies containing a bunch of elements that are nothing but ashes of star explosions. This sounds to me as first class poetry, as is also the beautiful title of Hoyle's obituary. However, in my humble opinion, to say that the mathematical formalism of inflation is just the same as that of Hoyle's coauthored Steady State model is simply going too far, it is not certain, as also many inflation specialists will tell you. What is indeed true, as I want to stress again, is that the {\em underlying physics}, the physics that allows creating matter out of negative pressure and keep the energy balance till the end of the process, is certainly the same physics of inflation, as beautifully explained in Tolman's book of 1934 \cite{tolm1}. But this is ultimately, in essence, just General Relativity, interpreted, as it should be, in the proper way.

An important remark is here in order, namely that Hoyle invoked the Big Bang (which you may call now `inflation') just with one single purpose, namely in order to create matter/energy, and not for any of the other crucial reasons which ultimately led to its formulation by Allan Guth in his very famous paper published on 15 January 1981 \cite{guth1}, namely the horizon problem, flatness, causality, the monopole problem, and so on. This should have been clear already from the above discussion, but it must be     properly stressed.

\section{The many different faucets of the concept of Big Bang}

The confusion we have addressed above, which arises in the definition of the term Big Bang in the popular literature, is even worsened because the same expression has been used, since 1949, in other different contexts. Until now, we have here only considered the original meaning of Big Bang, namely the one coming out of the mind and exact words of Fred Hoyle, the popularizer of such expression. But it turns out that in the almost seventy years elapsed since then, the same two words have been used in more or less related subjects, notably, in physics,  the Big Bang Singularity, the Hot and Cold Big Bang Models, and outside of physics in innumerable contexts, in novels, movies, and in a very popular TV series. No wonder, this last is actually the first meaning that inevitably appears now when one does a search on the internet. Thus, whenever one uses the term ``Big Bang'' one should immediately specify which of these concepts one is actually refereeing to. This is too often not the case in the literature, adding to the general confusion.

\subsection{The Big Bang Singularity}

Nowadays, in many scientific sources, including pictures and plots of the evolution of the Universe, the most common use of Big Bang is to refer to the singularity at the origin of the Universe. I will not discuss this second meaning of Big Bang in much detail, since this is not the main scope of this paper, but just give a brief summary.

\subsubsection{The Belinsky-Khalatnikov-Lifshitz and the Misner Singularities}

In the 1960s, one of the main cosmological issues being studied by the Landau group in Moscow was about the possible time singularity at the origin of the Universe. In particular: (i) whether cosmological models based on general relativity necessarily contain a time singularity, or (ii) if actually the time singularity was an artifact of the assumptions used to simplify these models (such as homogeneity and isotropy of the universe). In several papers published between 1963 and 1971 \cite{bkl1}, Belinsky, Khalatnikov and Lifshitz (BKL)  proved that the universe oscillates around a gravitational singularity, in which time and space become equal to zero.

They also showed that the singularity is not artificially created by the simplifications made by the other special solutions, such as the Friedmann-Lema\^{\i}tre-Robertson-Walker, quasi-isotropic, and Kasner solutions. Their model was described by an anisotropic, homogeneous, chaotic solution to Einstein's field equations of GR.

In 1969  Charles Misner constructed a similar model \cite{mis1}, named the mixmaster universe, which was also homogeneous but not isotropic, and which would expand in some directions and contract in others, with the directions of expansion and contraction changing repeatedly, what suggested that the evolution was in fact chaotic.

\subsubsection{The classical Singularity Theorems}

We will here summarize in a unified fashion just two of the main singularity theorems that appeared starting in the mid 1960's for Einstein's field equations without further specific assumptions, as homogeneity or isotropy. The starting point here was the pioneering Penrose singularity theorem \cite{pen1} of 1965  (for a few more references, see \cite{cst1,cst1b}). Roger Penrose closed in fact the loophole discussed above by showing that, under very general assumptions, the singularity is unavoidable. Penrose proof relies on the concept of incomplete geodesics.
\medskip

\noindent{\bf Theorem 1 (Big Bang).} {\em Let $(M,g)$ be a global hyperbolic spacetime satisfying $R_{ab} X^aX^b \geq 0$ (being $R_{ab}$ the curvature tensor) for all temporal vectors $X^a$   (Einstein's Eqs. with the strong energy condition.) If there exists a spatial Cauchy $C^2$ hypersurface, $S$, for which the trace of the intrinsic curvature $\kappa$ satisfies $\kappa<k<0$, with $k$ const., then no temporal curve starting from $S$ and going towards the past can have a length that is larger than $3/|k|$.}

{\em All temporal geodesics to the past are incomplete.}

This is to say, under the conditions observed to be true for our Universe (Hubble's law) and admitting the validity of General Relativity, our Universe did have a beginning, in a singularity, which everybody now calls {\em the Big Bang singularity}.
\medskip

\noindent{\bf Theorem 2 (Black Holes).} {\em Let $(M,g)$ be a global hyperbolic space-time satisfying $R_{ab} L^aL^b= 0$ (being $R_{ab}$ the curvature tensor)  for all lightlike vectors $L^a$ (Einstein's Eqs. with the strong or the weak energy conditions.) Assume that there is a spatial Cauchy $C^2$ hypersurface, $S$, and a trapped surface, and let $\tau_0$ be the maximum value of the expansion over it. If  $\tau_0 < 0$,  there exists at least a lightlike geodesic that cannot be extended to the future, and is orthogonal to the trapped surface. Moreover, the value of the affine parameter, up to the point of no further extension of the geodesic, is less than $2/|\tau_0|$.}

In other words, the existence of a non-extensible lightlike geodesic implies that there will be a lightlike observer (e.g. a photon), which, starting from that surface and after a time of travel proportional to $2/c|\tau_0|$, will necessarily fall into a future time singularity.

In any case, since we do not have  a theory of quantum gravity we cannot know with certainty the physical nature of this singularity.

The following important consideration is here in order. It turns out that, before one ever reaches the singularity at the origin of everything, classical physics, as described by Einstein's field equations, ceases to be valid. Actually, we need not go so far back in time in order to experience this: try to describe the Hydrogen atom with GR, this has simply no sense. Thus, in a way, the classical singularity theorems, although mathematically rigorous, lack physical meaning.

Those were the reasons of the Moscow Russian  school again, now however around Yakov Zel'dovich (Starobinsky, Mukhanov, Chibishov, ...) at the end of the 1970s, for invoking the inescapable necessity of quantum corrections to the gravity equations --a limited possibility but the only one available, given that a rigorous theory of quantum gravity was, and is still, lacking. It was during a visit of Hawking to Moscow in the early 1970s where he got the idea of adding quantum corrections to black hole physics, what resulted in his extremely important discovery of the Hawking radiation (as is now called), as Hawking himself has recognized several times.

\subsubsection{On the BGV (Borde-Guth-Vilenkin) Theorem}

Inflationary cosmological models seemed bound to invalidate the conditions of all the classical singularity theorems above \cite{guth1,infl1}. In the 1980's, it was attempted (without success, in fact) to construct models that, starting from an exact de Sitter  solutions would be past eternal.
In 1994 Borde and Vilenkin (BV94) proved an extended theorem \cite{bv94}, which states that inflationary spacetimes are past geodesically incomplete, what again implies, in other words, an initial singularity.
The key assumption for this theorem was that the energy-momentum tensor obeys the weak energy condition (WEC), what was an advance over the previous theorems.

However, quantum corrections to inflationary models seem to violate such condition, too, when quantum fluctuations result in an increase of the Hubble parameter: $dH/dt > 0 $, which, on the other hand, is an essential condition for chaotic inflation to be eternal.  Thus, the WEC must be generically violated in those models!   And this seemed to open the door to a scape from the BV theorem.

Such was the motivation for Borde, Guth \& Vilenkin in his celebrated paper  \cite{bgv1}  	{\em ``Inflationary Spacetimes Are Incomplete in Past Directions''}.  As the title already indicates, they recovered again the result of BV94, but with a few important additional considerations, which have not been fully appreciated by some inconditional supporters of the creationists theories. I will not further discuss this issue here \cite{bgvcr1}.

Technically, one starts now from the (sufficient) condition of a ``quasi dS'' state with a minimal condition of  ``averaged expansion'' (along $t$-paths):        $$H_av > 0,$$
that is, an average taken over all time trajectories.

Moreover, the theorem can be extended to extra dimensions, and also to cyclic models \cite{st02}, for $H_av > 0$ in those models.

As a consequence, in all these cases, and under the strict conditions of the theorem, one gets initial geodesic incompleteness! That is, an origin, once again \cite{bgv1}.\footnote{In a way, one could say that inflationary spacetimes are as singular as the steady state universe, what maybe can be brought back to the partial incompleteness of de Sitter spacetime \cite{jbarrpc1}, already noted in \cite{cst1} and also discussed in \cite{barrT1}.}

\subsection{A quick sketch of the possible origin of the Universe}

A rather extended view today is that the origin of the Universe took place out of nothing --or almost nothing. Let us be more precise (for a more detailed account of the possible origin of the universe `from nothing', see e.g. \cite{krau12}; for some scientific critical comments, see \cite{kohli14}.)

The first question that occurs to us is: What is {\em nothing}? The answer of modern physics is (at the very least) twofold, namely
\begin{enumerate}
\item In fundamental classical physics the ultimate theory is GR, and there the vacuum solution is the de Sitter solution (the zero-energy one) of Einstein's field equations.
\item In quantum physics, on the other hand  `nothing' means
	 the vacuum state of the quantum system at hand, e.g., in our case the one at the very beginning of all, as far as we can go into the past.
\item It needs little explanation that we are missing here the actual theory that we would need in order to answer the question with more property, namely the theory of quantum gravity (QG).
\item However, it is not clear at all that, even in possession of QG, we would be allowed to penetrate the Planck domain, which establishes a limit to all known theories of Physics (quite probably also to this unknown QG).
\end{enumerate}
And this is the state of the art of such fundamental issue today. Letting aside the Planckian constraint and ensuing considerations, some possibilities have been recently proposed with essentially two different, so to say, minimalist starting  points:
\begin{enumerate}
\item Just quantum spacetime, and nothing else! In spite of some attempts to do that (as most recently by Lawrence Krauss and Frank Wilczek) nobody has convincingly succeeded yet.
\item In addition, a scalar field Hamiltonian (or two), namely the Higgs, an inflaton,... This seems of course a more feasible possibility, at the expense of having to explain where do these additional necessary fields come from.
\end{enumerate}

\subsection{Big Bang Cosmological Models}

Just a rather brief mention, accompanied with a number of basic references, of this vast subject, which was at the heart of the most fundamental discussions about cosmology for some generation of physicists, namely hot or cold Big Bang model? clearly decided in favor of the first. A question that had an even more important second part, under the form of hot or cold dark matter? eventually decided in favor of the second.

\subsubsection{The Hot Big Bang Model}

Skipping now the details of the Big Bang and previous to the formulation of inflation, with its many specific theories, what remained true of the  original idea of Lema\^{\i}tre was that, in fact, in the past the Universe was in a very dense state and very hot, so that matter could relax to statistical thermal equilibrium. Black body radiation was filling the whole space. With the expansion of the Universe the temperature went down until the first (neutral) atoms could form, namely Hydrogen ones, and radiation could travel across the whole Universe under the form of what is now called the CMB or cosmic microwave background radiation (formerly also called CBR, cosmic background radiation). Some useful references where this process is described in detail are \cite{calca1}-\cite{Bwww7}.

In the beautiful description of Lema\^{\i}tre, the ulterior processes which took place can be compared with fireworks:
 \begin{quotation}
{\em ``The evolution of the world can be compared to a display of fireworks that has just ended; some few red wisps, ashes and smoke. Standing on a cooled cinder, we see the slow fading of the suns, and we try to recall the vanishing brilliance of the origin of the worlds.''}
\end{quotation}
In the time interval from about two to thirty minutes, but mostly within the first three minutes after the Big Bang (see \cite{web1} for a very popular reference), an efficient synthesis of the light elements, namely Deuterium, Helium-3, and Helium-4 took place. This is what is called the era of primordial nucleosynthesis. The current abundances of light elements are in accordance with what happened during that time, placing strong constraints on the state of the Universe then, and particularly on the baryon density. Our Universe contains now some 23\% of its mass in Helium (its production in stars is not relevant, as compared to the primordial production during the first three minutes). The conditions then had to be precisely those leading to our Universe, which has nine Hydrogen nuclei for every Helium nucleus \cite{Bwww2}. Moreover, it is well known now that most of the Universe's Hydrogen is in its simplest form and  not in heavier isotopes, namely deuterium or tritium. Deuterium, on its turn, is not produced, but only destroyed, in stars, so that its abundance today sets a lower limit on the amount of deuterium from primordial nucleosynthesis, and again on the density of baryons, too.

The Hot Big Bang model does explain what we see in our Universe. To summarize further evidence, we know list what have been called sometimes the four pillars of the standard Hot Big Bang model: \cite{Bwww1}
\begin{enumerate}
	\item The  Universe expansion.
\item The origin of the CMB.
\item The primordial nucleosynthesis of the light elements.
\item The formation of the galaxies and of large-scale structures.
\end{enumerate}

After some six decades of dealing with this model with considerable success, a crisis took place at the beginning of the 1990s \cite{Peebles91}, just anticipating the discovery of the acceleration of the Universe expansion, which completely changed the paradigm \cite{Khoury12} and, in particular, the energy content of the Universe. Some important consequences of this discovery, in special concerning the possible appearance of finite time future singularities, will be discussed in the next Section.

\subsubsection{The Cold Big Bang and other models}

The idea of a possible Cold Big Bang goes back to Lema\^{\i}tre's theory of a primeval atom, forming a gigantic ball of nuclear liquid in a state at very low temperature, which was required in order to keep it from falling apart via thermal fluctuations. In Lema\^{\i}tre's words \cite{alberg1}:
\begin{quotation}
	{\em ``If matter existed as a single atomic nucleus, it makes no sense to speak of space and time in connection with this atom. Space and time are statistical notions that apply to an assembly of a great number of individual elements; they were meaningless notions, therefore, at the instant of first disintegration of the primeval atom.''}
\end{quotation}
Ultimately, this idea was not able to explain the Universe expansion and the origin of the light elements.

A variant on Lema\^{\i}tre's cosmology was proposed in 1966 by David Layzer \cite{infil1}, who developed a short-lived alternate to the standard Hot Big Bang cosmology by proposing that the initial state was near absolute zero, thus reminiscent of Lema\^{\i}tre's initial state. Through thermodynamic arguments, Layzer argued that rather than the universe starting in a high entropy state, it began with a very low entropy (see also \cite{barbo1}). Anyhow, the CMB radiation is very difficult to explain in these theories, in spite of some more recent attempts \cite{agui1}. To finish, most of the  versions of a Cold Big Bang being considered  predicted an absence of acoustic peaks in the cosmic microwave background radiation, a possibility that was eventually  ruled out quite clearly by WMAP and, most recently, PLANCK observations.

For completeness, some other theories, which are in a way alternative to the Hot Big Bang model, can be found here \cite{wett1}-\cite{pbs1}.

\section{Acceleration: new singularities}

According to the most recent and accurate astronomical observations, it is very likely that our universe had an origin, some 13.8 billion years ago, from nothing (or almost nothing, a quasi de Sitter space) --e.g., from a vacuum state of a tiny quantum system including space-time and possibly a scalar field--  and is currently in accelerated expansion. In many of the models of modified gravity, which have been discussed in order to obtain this accelerated expansion new singularities, in a way similar to the Big Bang one, have shown up.

Recall the second Friedmann equation
\begin{equation}
\frac{\ddot{a}}{a} = -\frac{4 \pi G}{3} \left(\rho + \frac{3p}{c^2} \right)+ \frac{\Lambda c^2}{3},
\end{equation}
where $a$ is the scale factor, while $\rho$  is the matter/energy density, $p$ the pressure, and $\Lambda$ the cosmological constant.

For a fluid with equation of state     $P = w \, \rho$, where $w$ is the so-called equation of state parameter, the following three possibilities appear, according to the different values this fundamental parameter $w$ may have, and which we already know, from the most recent and accurate astronomical observations, to be $w \sim -1$. Namely,
\begin{itemize}
\item $w=-1$, {\bf  the cosmological constant case.} The simplest and most natural in general relativity, but difficult to explain, and it seems to need for a symmetry no one has been able to find yet, in a convincing way, in order to solve the associated cosmological constant problem.
\item $w>-1$, {\bf  so-called quintessence case.} It is the most ordinary case and usually involves an evolving scalar field.
\item $w<-1$,  {\bf the phantom case.} It involves a so-called phantom field (of negative kinetic energy) and leads to a number of future singularities at finite (or infinite) time.
\end{itemize}

If the universe is now, indeed, in the $\Lambda$CDM era (that is, cold dark matter with a cosmological constant $\Lambda$), it might remain in such era, eventually becoming asymptotically de Sitter, i.e., a regular universe during all its future evolution. This would be the most simple and natural situation, were it not for the annoying cosmological constant problem.

For a phantom or quintessence dark energy era, other singularities appear (for a few seminal references, see \cite{Caldwell:1999ew,Caldwell:2003vq,Barrow:2004xh,Barrow:2004hk,BouhmadiLopez:2006fu}). According to Ref.~\cite{Nojiri:2005sx}, they can be most conveniently classified as follows.
\begin{itemize}
\item {\bf  The Big Rip}  or {\bf Type I} singularity  (occurring in a phantom dominated universe) \cite{Caldwell:1999ew}. In the limit  $t = t_{s}$ (a finite value of time in the future) all quantities, namely the scale factor, effective energy density and pressure of the universe diverge.
\item {\bf  Sudden Singularity}, or {\bf  Type II}, discovered in \cite{Barrow:2004xh}\footnote{See also \cite{BarrowGT86}, where this type of  sudden and other finite time singularities were first introduced in order to show that closed Friedmann universes need not collapse even if 	they satify the strong energy condition; although the terminology `sudden singularity' was introduced by Barrow in 2004, as already mentioned.}. In the limit  $t = t_{s}$  only the effective pressure of the universe becomes infinite, while the scale factor $a$ and the total effective energy density $\rho$ remain finite.
\item {\bf  Type III} or {\bf  Big Freeze} future singularity. In this case, only the scale factor remains finite, while  both the effective energy density and pressure of the universe diverge at $t = t_{s}$. These can be either weak or strong singularities, which are geodesically complete solutions.
\item {\bf  Type IV} or {\bf  Generalized Sudden} singularity. In the limit $t = t_{s}$ none of these, the scale factor, effective energy density or pressure diverge. However, higher derivatives of the Hubble rate $H$ become divergent (not $H$ itself and its first derivative), as discovered in \cite{Nojiri:2005sx}, where a full classification was given. In this case a weak singularity appears and geodesics are complete.
\end{itemize}
Eventually, the universe may survive the passage through a Type IV singularity or a Sudden Singularity (Type II).
\begin{itemize}
	\item And there is still the case of the so-called {\bf  Little Rip} universes, where the future singularities occur asymptotically, at infinite time only.
Typically, this happens when the scale factor increases rapidly, as $a(t) = \exp[\exp(t)]$ or higher exponentials \cite{fls1,brev1}. But different combinations are also possible, as an oscillating universe (bounce). Also, the very important point must be remarked  that quantum gravity effects may affect this future evolution, preventing a Big Rip to occur by such quantum effects \cite{Elizalde:2004mq} or by a similar Casimir-type effect or other (see, e.g., \cite{other1}).
\end{itemize}
Some new singularities have been discovered recently to be added as an update to this table:
\begin{itemize}
	\item {\bf  The $w$-singularity}, also called Type V, characterized by the fact that only the barotropic index $w$ is divergent while the other quantities are smooth \cite{other1a}.
	\item {\bf The $Q$-singularity}, which appears in models where there is an interaction between dark matter and dark energy. On top of the above future singularities, a divergence of the time derivative of $w$ may give rise to a singular interaction. In fact, perturbations may be sensitive to it, since the adiabatic sound speed in a barotropic fluid depends on $w'$ \cite{other1b}.
\end{itemize}
Increasing effort is being devoted towards the classification of singularities in all sort of cosmological models. In particular, there is an ongoing intensive study of cosmic singularities for very different modified gravity theories. More results and information related to the above tables can be found in  \cite{dabd1} (see also the references therein).

\section{Conclusions}

As explained in the first section of this paper, there are powerful reasons to consider 1912  as the Year that  marks the beginning of modern cosmology. The author is now fully convinced that it should be officially declared as such. Indeed, two extraordinary, almost simultaneous developments occurred that Year, to wit: (i) the publication by Henrietta S. Leavitt of her crucial law, the period-luminosity relationship of Cepheid variable stars, namely a linear dependence of the luminosity vs the logarithm of the period of variability, and (ii) the beginning of Vesto Slipher's investigation of the velocities of spiral nebulae, obtained by means of their corresponding spectral deviations, as red- (or blue-) shifts. These two extremely powerful tools allowed, in the few following years, the formulation of Hubble's law and its matching (by Georges Lema\^{\i}tre) with the expanding solution of the Einstein field equations, a solution that actually had been first obtained by Alexander Friedmann, a couple of years before. The Universe was expanding, indeed! It took some time, even to astronomers and theoretical cosmologists (in Einstein's case four full years) to understand this very difficult concept. In many cases it took even decades, what proves that this one is far from being an easy notion, as we have discussed in the paper in sufficient detail.

The fine tools put in place by Leavitt and Slipher, in 1912, where used by Hubble and many other astronomers subsequently --as were also, in parallel, the Einstein field equations, from 1915,  by Friedmann, de Sitter, Lema\^{\i}tre and so on-- to craft the extraordinary structure, still under construction, of Modern Cosmology. Therefore the proposal of the Year 1912 as the one marking the birth of this discipline.

The second part of the paper has been devoted basically to the history of the original meaning of the Big Bang concept, a definition that is too badly explained in the available generalist literature. Fred Hoyle had a very clear idea in mind when he said these two words emphatically in 1949. Indeed, the complete sentence he pronounced on that occasion is perfectly understandable and absolutely meaningful to anyone who is versed in General Relativity, but not so at all to non-specialists, and this is maybe the main reason for the unbelievable confusion generated around this term. Another one being that the same two words, namely `Big Bang', have been subsequently used in a number of different situations, most notably to give name to cosmological models, e.g. the hot and the cold Big Bang models, and to the Big Bang singularity at the very origin of space and time. In this paper I have defended the thesis that all this new terminology has practically buried the original concept, which we have had here to rescue from its ashes, with the help of recent important discoveries from historians of science. Needless to say, all the above is quite well known to professional cosmologists and astrophysicists with a solid background, but not so, regretfully, to many other scientists and intellectuals, in general.

As a final conclusion, owing to the fact that these two magic words, Big Bang, have become so extremely popular in many different contexts and mass media channels (even little children use them sometimes), all of us, scientists who know about this matter, must consider as our inescapable duty a serious compromise to explain the precise meaning of Big Bang in a fair, understandable, but non-misleading way, not only to students, at our universities, institutes, and schools, but also to the general public, anywhere and at any given opportunity.


\bigskip

\noindent {\bf Acknowledgements}. This work was partially supported by CSIC,
	I-LINK1019 Project, by MINECO (Spain), Projects FIS2013-44881-P and
	FIS2016-76363-P, by the CPAN Consolider Ingenio Project, and  by JSPS (Japan), Fellowship N. S17017. This research was started while the author was visiting the Kobayashi-Maskawa Institute
	in Nagoya, Japan. He is much obliged with S. Nojiri and the rest of the members
	of the KMI for very warm hospitality. The paper was finished during a research stay at Dartmouth College, NH, USA. The author thanks the anonymous referees of the paper for interesting remarks. Finally, he thanks John Barrow, Mariusz Dabrowski, Diego S\'{a}ez-G\'{o}mez, and Andrzej Wr\'{o}blewski  for important comments that have been incorporated to this new version.

\end{document}